# The Real Interest Rate as a Control Variable in the Open Economy

Carlos Esteban POSADA P.[1]

Liz LONDOÑO-SIERRA.[2],[3]

*Abstract*

*This paper addresses the structure and dynamics of an open market economy and its relations with the real interest rate. In this respect, the paper is situated within a broad conventional literature. However, it departs from the standard approach to the interest rate by treating it as a control variable. Even so, the analysis concludes that the two main determinants of the interest rate are the future utility´s discount rate and expectations regarding future multifactor productivity (labor efficiency). Furthermore, increases in such expectations lead to increases in both the interest rate and wages. These results are consistent with those obtained with the Cass-Koopmans-Ramsey model.*

## ***1. Introduction***

In this paper we address in theoretical terms the topic of the real interest rate and its relations with the structure and dynamics of an open market economy by presenting and applying a two-period model without reference to money, nominal magnitudes or monetary policy. In many respects the article falls within the mainstream literature. However, it departs from the standard approach by demonstrating that the real interest rate is determined by the economy´s structural characteristics and expectations on multifactor productivity (labor efficiency) and is also a control variable.

The remainder of the paper is organized as follows. Section 2 reviews the related literature. Section 3 presents the model. Section 4 (*Discussion*) makes a statement on our approach and conceptual framework. Section 5 develops the intuition behind the model's implications for the capital flows and labor markets. Section 6 summarizes and concludes. In the Annex we show numerical results of comparative statics exercises with the model.

## ***2. Related Literature***

The analysis of the real interest rate in open economies is part of a broad theoretical tradition in International Macroeconomics that can be traced back to the seminal contribution of Metzler (1960), who formulated a framework in which the international real interest rate equilibrates global saving and investment thereby generating creditor and debtor positions across countries. In this framework, the interest rate acts as the intertemporal price that coordinates consumption, saving and investment decisions at the international level. The version developed by Obstfeld and Rogoff (1996) provides explicit microeconomic foundations and considers two periods, present and future, and two economies, one in trade account´s surplus and one in deficit. The former finances the latter through credit. The interest rate is international, and each economy takes it as exogenous. This framework became the standard reference in modern International Macroeconomics, in which small open economies take the international interest rate as given and external borrowing emerges as the mechanism that allows consumption to be smoothed over time.

[1] Professor; School of Finances, Economics and Government; Universidad EAFIT. Address: cposad25@eafit.edu.co

[2] Part Time Professor; School of Finances, Economics and Government; Universidad EAFIT. Address: llondo11@eafit.edu.co

[3] We thanks José Miguel ARIAS, Carlos Andrés BALLESTEROS and Cristian CASTRILLÓN for their comments and help.

On the other hand, the neoclassical growth tradition, particularly related to Cass-Koopmans-Ramsey model (Wickens, 2011), Barro (1974, 1989) and Barro and Sala-i-Martin (2004), stablishes the equilibrium real interest rate as a variable determined by fundamental real factors[4].

In addition, it is common in the conventional literature, following Metzler, to define the international real rate as one free of default´s risk, and to assume some countries are borrowing at that rate while others cannot. The latter may still access to external financing but at an interest rate equal to the international rate plus a spread associated with lenders' assessment of such a risk. Eaton and Gersovitz (1981) developed the fundamental theoretical framework for analyzing sovereign borrowing under default´s risk, showing spreads reflecting both the probability of default and the losses (costs) associated with debt renegotiation. Building on this foundation, Arellano (2008) presented a sovereign default model in which the risk premium emerges endogenously as a function of the level of indebtedness and the state of the economy. This approach has been central to explaining the dynamics of sovereign spreads in emerging economies.

The theoretical literature on default´s risk is extensive and varied. However, in macroeconomic models designed for numerical exercises or to support econometric estimation, it is common to introduce the spread in an *ad hoc* manner by assuming that its magnitude depends on the size of debt, as in Uribe and Yue (2006) and Uribe and Schmitt-Grohé (2017). Aguiar and Gopinath (2006) and Neumeyer and Perri (2005) provide empirical evidence on the importance of country risk premium shocks as a source of macroeconomic fluctuations in emerging economies, showing that increases in the spread can generate deep recessions through the credit cost channel. Reinhart and Rogoff (2009) show that episodes of indebtedness, crisis, and default are recurrent, reinforcing the relevance of this mechanism for explaining the observed dynamics of interest rates in emerging markets. In addition, the conventional literature recognizes the defaults may occur despite previously agreed risk premia if unexpected shocks arise making it optimal for debtors to repudiate their commitments.

[4] Del Negro *et al*. (2019) found econometric evidence for the two determinants of their long-term global real interest rate indicator (constructed using yields on "safe and liquid" securities from the United States, the United Kingdom, Japan, Germany, France, Canada, and Italy between 1870 and 2016): a) "willingness to pay for such securities" (a "propensity" impossible to consider in models without uncertainty), and b) the rate of economic growth. The authors conclude that the trend decline in the real interest rate is, fundamentally, a phenomenon from 1980 onwards and is caused by an increase in the aforementioned "willingness to pay for…," and by the decline in the rate of economic growth.

The literature has extensively examined the conditions under which interest rate parity holds. Obstfeld and Rogoff (1996) provide a comprehensive analysis of covered and uncovered interest rate parity, establishing that, in the absence of frictions and under perfect capital mobility, domestic nominal interest rates must equal to the international rate adjusted for expected depreciation. In other words, the country takes the international interest rate as given, and domestic saving does not need to equal domestic investment, since the difference is financed through international capital flows. However, empirical evidence suggests that this result holds only imperfectly. Barro *et al*. (1995) find mixed evidence on international convergence in real interest rates, which they interpret as indicative of financial frictions and market segmentation. These findings have motivated the development of models relaxing the assumption of perfect financial integration. Schmitt-Grohé and Uribe (2003) explicitly recognize that this assumption generates indeterminacy in models of small open economies, requiring the imposition of stationarity conditions such as portfolio adjustment costs or debt-dependent risk premiums.

The literature also includes models with financial frictions leading to the determination of (endogenous) interest rates even in small open economies. In this context, the external finance premium depends on borrowers' net worth in the presence of asymmetric information. In this environment, the effective real interest rate relevant for private optimization is determined endogenously, although a risk-free exogenous benchmark rate remains. In the international context, Céspedes *et al*. (2004) show that these frictions can amplify adverse shocks in emerging economies, especially when borrowing is denominated in foreign currency. While these models allow for endogenous variation in the cost of credit, they preserve the fundamental distinction between an exogenous base rate and an endogenous premium. Chari *et al*. (2005) develop models in which spreads do not arise *ad hoc* but emerge from specific informational frictions, allowing for an analysis of how domestic conditions affect the cost of external borrowing. Mendoza (2010) proposes models in which financial crises can arise even in the absence of exogenous shocks through endogenous amplification mechanisms, suggesting that the interest rate relevant for the economy cannot be treated as an exogenous variable.

From a methodological perspective, it is important to highlight the tradition of deterministic perfect foresight models in Macroeconomics. Barro (1979) uses this approach to analyze public debt and Ricardian equivalence in a multigenerational setting. Bohn (1995, 1998) shows that if fiscal policy responds positively to the level of public debt by increasing surpluses the public debt´s path remains sustainable and default does not occur. In the context of open economies, Schmitt-Grohé and Uribe (2003) develop deterministic models to analyze external debt sustainability, showing that even in the absence of uncertainty multiple equilibria may arise depending on expectations about future fiscal policy. These models allow for a precise characterization of equilibrium trajectories and are particularly useful for analyzing permanent changes in policies. The absence of uncertainty eliminates the possibility of default which represents both a limitation, as it does not capture debt crises, and a strength, as it allows for the fundamental determinants of the equilibrium interest rate to be isolated without interference from stochastic shocks. Végh (2013) makes extensive use of deterministic models to analyze fiscal policy and the current account in open economies.

Can the existence of an exogenous international interest rate be omitted in a model? An affirmative answer has precedents in the literature. Backus *et al*. (1992) illustrate this one through a model with two economies, each with an infinite horizon, in which cyclical movements include fluctuations in external deficits and surpluses. When one economy runs a deficit it demands credit, while the other, running a surplus, supplies it. In steady state, the real interest rate is common to both economies and equal to the subjective discount rate of future utility. Frenkel and Razin (1992) present a two-period, two-economy model with a single real interest rate (the international rate) which is based on the discount rates of future utility and the growth rates of output in both economies. Kehoe and Levine (1993) develop models in which borrowing limits arise endogenously from the condition that agents do not find it optimal to repudiate their debt. Álvarez and Jermann (2000) show that such constraints can significantly limit capital flows across countries even in the absence of explicit informational frictions. Antràs and Kulesza (2026) present a two-economy model: the national and the "rest of the world" economies. Under certain conditions (and without uncertainty) the interest rate in the national economy may differ from that of the rest of the world, and both rates are determined endogenously by their own productive and financial conditions, as well as by their levels of impatience.

## 3. *The Model*

### *3.1. The Household´s Problem*

The consumer (the representative household´s head) maximizes the household´s welfare, $U$, where it is defined as the weighted sum of utility in the present and future periods (periods 0 and 1):

$$(3.1.1)\quad U = u_0 + \beta u_1;\ \beta \equiv \frac{1}{1+\rho};\ \rho > 0$$

Where $u, \rho$, 0, 1 denote the period utility, the subjective discount rate of future utility, and the indicators of the present and future periods, respectively.

The welfare maximization is subject to the intertemporal budget constraint and to the time constraint (measured in hours) available for work. These constraints are given by:

$$(3.1.2)\ \ c_0 + \frac{c_1}{1+r} \leq w_0 l_0 + \frac{w_1 l_1}{1+r} + x_0 + \frac{x_1}{1+r} - tax_0 - \frac{tax_1}{1+r}$$

$$(3.1.3)\ \ 0 < l_0 \leq l_0^{max}; 0 < l_1 \leq l_1^{max}$$

Where $c, r, w, l, x, tax$ denote household consumption, the real interest rate, the hourly wage, the number of hours worked, the dividend (participation in the firm's profit), and the total tax paid by the household, respectively.

The following utility function is assumed, namely the *KPR* function (King, Plosser, and Rebelo), separable in consumption and labor:

$$u = \frac{c^{1-\gamma}}{1-\gamma} - \varphi \frac{l^{1+\theta}}{1+\theta};\ \ 0 < \gamma, \theta, \varphi$$

Where $\gamma$ is the inverse of the of consumption´s substitution intertemporal elasticity, $\varphi$ is the weight of the of labor´s disutility in total utility, and $\theta$ is the inverse of the elasticity of labor supply with respect to the wage (Frisch elasticity).

From this point onward, and throughout the remainder of the document, it is assumed that the number of households in the present and in the future, $N_0, N_1$, is equal to the number of consumers and to the number of suppliers of labor services.

The maximization of (3.1.1), subject to (3.1.2) and (3.1.3), and given the utility function specified above, is summarized by the first-order conditions. These, in turn, generate the following equations for present and future consumption, and for current labor supply, assuming (for simplicity) that future labor supply is an exogenous variable:

$$(3.1.4)\ c_1 = [\beta(1+r)]^{\frac{1}{\gamma}} c_0$$

$$(3.1.5)\ c_0 = \left(\frac{1}{Q}\right)\left(w_0 l_0 + \frac{w_1 l_1}{1+r} + x_0 + \frac{x_1}{1+r} - tax_0 - \frac{tax_1}{1+r}\right)$$

$$(3.1.6)\ Q \equiv 1 + [\beta(1+r)]^{\frac{1}{\gamma}}\left(\frac{1}{1+r}\right)$$

$$(3.1.7)\ l_0 = \left[\frac{\beta w_0}{\frac{w_1}{1+r}}\right]^{\frac{1}{\theta}} l_1$$

***3.2. The Firm´s problem***

Since there are only two periods, investment takes place in the first period (gross investment: $I_0 > 0$), while in the second period investment is zero (it is assumed that the capital depreciation rate is equal to 1. Thus, the capital stock at the end of the second period is zero[5]).

The firm's problem[6] and its solution are summarized by the first-order conditions of optimization which, given the assumptions of positive but diminishing marginal productivities of labor and capital, are also sufficient conditions. These conditions are expressed as equations (3.3.7), (3.3.8) and (3.3.9) of the model presented in the following subsection.

### *3.3. The Macroeconomic Model*

$$(3.3.1\ and\ 2)\ \ Y_i = \ K_i^{\alpha}(A_i L_i)^{1-\alpha};\, i = 0{,}1;\ 0 < \alpha < 1$$

$$(3.3.3\ and\ 4)\qquad L_i \equiv N_i l_i$$

$$(3.3.5)\ \ l_0 = \left[\frac{\beta w_0}{\frac{w_1}{1+r}}\right]^{\frac{1}{\theta}} l_1 \ \leq l_o^{max}$$

$$(3.3.6)\ \ l_1 = l_1^{max}$$

$$(3.3.7\ and\ 8)\ \ w_i = (1-\alpha)\frac{Y_i}{L_i} = \frac{\partial Y_i}{\partial L_i}$$

$$r = \alpha\frac{Y_1}{K_1} - \delta = \frac{\partial Y_1}{\partial K_1} - \delta \ \ \Rightarrow$$

$$(3.3.9)\ \ K_1 = A_1 L_1 \left(\frac{\alpha}{\delta + r}\right)^{\frac{1}{1-\alpha}}$$

$$(3.3.10)\ \ I_0 = K_1 - (1-\delta)K_0$$

$$(3.3.11)\ \ c_0 = \left(\frac{1}{Q}\right)\left(w_0 l_0 + \frac{w_1 l_1}{1+r} + x_0 + \frac{x_1}{1+r} - tax_0 - \frac{tax_1}{1+r}\right)$$

$$(3.3.12)\ \ Q \equiv 1 + [\beta(1+r)]^{\frac{1}{\gamma}}\left(\frac{1}{1+r}\right);\ \beta \equiv \frac{1}{1+\rho}$$

$$(3.3.13)\ x_0 = \frac{1}{N_0}(Y_0 - w_0 L_0 - I_0)$$

$$(3.3.14)\ \ x_1 = \frac{1}{N_1}(Y_1 - w_1 L_1)$$

---

[5] There is, however, an alternative specification with similar macroeconomic results: assuming a depreciation rate lower than 1 and considering that the remaining capital at the end of the second period is transferred to households as an "extra dividend," which is then used for consumption.

[6] It is assumed that there is only one firm in the economy; this relies on the hypothesis of constant returns to scale.

$$(3.3.15)\ \ c_1 = c_0[\beta(1+r)]^{\frac{1}{\gamma}}$$

$$(3.3.16\ and\ 17)\quad C_i \equiv N_i c_i$$

$$(3.3.18)\ \ T_1 = (1+r)G_0 + G_1 - T_0(1+r)$$

$$(3.3.19\ and\ 20)\quad tax_i \equiv \frac{1}{N_i} T_i$$

$$(3.3.21)\qquad X_0 - M_0 = Y_0 - C_0 - I_0 - G_0$$

$$(3.3.22)\qquad X_1 - M_1 = Y_1 - C_1 - G_1$$

$$(3.3.23)\quad X_0 - M_0 = \frac{1}{1+r}(M_1 - X_1)$$

Where $Y, K, A, \delta, \alpha, I, T, G, X, M$ denote output, capital, labor efficiency, capital depreciation rate, output elasticity with respect to capital, gross investment, total tax revenue, government expenditure (public purchases of output), exports, and imports, respectively.

### *4. Discussion: Our Approach and Conceptual Framework*

The section 2 allows us to identify a theoretical gap. Although the open economy literature includes models with domestic endogenous real interest rates, there is, to the best of our knowledge, no model treating the real interest rate as a control variable in the specific context of an open economy. The model developed in this paper fills this gap.

A clarification is necessary at this point. The existence of an interest rate presupposes the participation of at least two agents in a credit transaction: one demanding credit and another one offering it. This paper focuses on the behavior and outcomes of one of these agents, the one representing the debtor or creditor of the economy under analysis, who supposedly negotiates with the agent representing the rest of the world.

Notwithstanding, to better understand the general meaning of the model in section 3, we consider the following representation of the world economy:

Let us assume the existence of three economies, each represented by a two-period model (present and future). If there are three economies, A, B and C could be three financial deals: A with B; A with C, and B with C, and three rates: $r^{AB}, r^{AC}, r^{BC}$, and it is possible that:

$$r^{AB} \lesseqgtr r^{AC} \lesseqgtr r^{BC}$$

The model of the section 3 could apply for each one of the three deals because each deal is independent of the other two. This differs from the models of Backus et al. (1992) and Frenkel and Razin (1992). In these two models there is only two economies, so there is one deal and one rate, and that deal is a deal for all the world.

Does the above mean that if $r^{AB} \neq r^{AC} \neq r^{BC}$ there is no free international capital flows? The simplest answer is yes: There is no free movement of capital if by this we understand that, given some initial differences in interest rates, global capital will be redistributed in such a way that achieving this redistribution will ultimately lead to the equalization of interest rates. And the model in section 3 does not assume that there are artificial obstacles to capital flows or risks of debt default; in this sense it can be stated that there is freedom of movement of capital but these do not necessarily lead to the equalization of the marginal productivities of capital (net of depreciation) and of the interest rates. The direction of the flows follows another criterion: that of making zero the present value of the sum of the deficits and surpluses of the foreign trade accounts of any one of the economies A, B and C.

But it is correct to assume that if there is freedom in the movement of capital flows, the negotiation of the interest rate $r^{BC}$ between economies B and C will take into account information about the rates $r^{AB}$ and $r^{AC}$. These latter rates will help establish the opportunity costs of what is negotiated between B and C and will undoubtedly influence the determination of the rate $r^{BC}$.

Let us now assume that, in the present period, economy A is in equilibrium between its exports and imports; therefore, there would only be one (international) negotiation of an interest rate between the agents of economies B and C. One agent would be the monopsonist (the one from the economy that demands credit because it has a trade deficit) and the other the monopolist (the one that offers credit because it represents the economy in the opposite situation). The negotiation of the interest rate would be indeterminate according to the theory of bilateral monopoly (see, for example, Ferguson, 1969, p. 281). But if both agents agree that the interest rate will be a control variable for both economies, there would be no indeterminacy. Furthermore, suppose that economy A does not exist; in this case, the only interest rate is the one determined in the negotiation between B and C. So, in the present period, one economy has a trade deficit and the other a surplus, and in the future the opposite will occur, with a common interest rate for both. This would eliminate any potential concern about whether or not there is a tendency for national interest rates to equalize.

### 5. *Investment and Saving, and Results from Exercises with the Model*

The similarities with the "conventional model" of an open economy, that is, with what would correspond to the prototype of the models referred to in the literature review, are many. The difference is the following: in the conventional open economy model the real interest rate is not a control variable, and it would be dissociated from the current account of the balance of payments in this same period or in any nearby period, whether that period corresponds to the steady-state path or to a transition phase. And such a dissociation is reasonable when the model has many periods.

By contrast, a two-period model with an endogenous real interest rate that is also a control variable requires establishing relationships between the real interest rate, investment, national saving, and the balance of the external current account in each of the two periods.

The advantage of taking this into account is that our model may be more useful to the analyst or policymaker who pays attention to economic policy issues associated with the determinants or the effects of the external current account or the labor market in the present period than a model, also neoclassical, in which the real interest rate is not a variable control.

We now seek to clarify the notions of saving, investment, and the real interest rate as implied by our model. For this purpose, it is sufficient to consider that:

$$(3.3.21)\ in\ (3.3.23)\ \Rightarrow$$
$$Y_0 - C_0 - G_0 - I_0 = \frac{M_1 - X_1}{1 + r}\ \Rightarrow$$
$$Y_0 - C_0 - G_0 + \frac{X_1 - M_1}{1 + r} = I_0$$

And with the following definitions:

$$S_0^N \equiv Y_0 - C_0 - G_0;\ \ S_1^x \equiv \frac{X_1 - M_1}{1 + r}$$

It turns out the following:

$$(5.2.1)\ \ S_0^N + S_1^x = I_0$$

In words, investment ($I_0$, recalling that there is investment only in the present period) is financed by present national saving ($S_0^N$) and "external saving" ($S_1^x$), the latter is defined as the present value of the future balance of the external current account. That is, if there is a trade deficit today, it is financed today by an inflow of foreign capital, but this inflow is exactly equal to the present value of the future surplus; therefore, it may be said that current investment is financed by current domestic saving and/or by future domestic saving.

Moreover, we know that:

$$I_0 > 0 \Rightarrow S_0^N + S_1^x > 0$$

But it is possible (depending on the values of the parameters and exogenous variables) that:

$$S_0^N \leq 0\ \Rightarrow S_1^x > 0$$

And also, it is possible that:

$$S_1^x < 0 \Longrightarrow S_0^N > 0$$

It should be clarified, however, that what we have called external saving ($S_1^x$), that is, the present value of the future surplus (deficit) of the external current account, decreases (increases) in response to increases (decreases) in the real interest rate, as follows:

$$\Delta^+ r \rightarrow \Delta^+(X_0 - M_0) \rightarrow \Delta^-(X_1 - M_1) \Rightarrow \Delta^- \left(\frac{X_1 - M_1}{1 + r}\right)$$

Therefore, equation (5.2.1) can be rewritten as:

$$(5.2.2)\ \ (S_0^N(r^+, \dots) + S_1^x(r^-, \dots) = I_0(r^-)$$

The possibility of an equilibrium of the model, considered as a whole (the model presented in Section 3.3), and that such equilibrium be at least locally stable, has, among its several necessary conditions, that the sum on the left-hand side of equation 5.2.2 have positive slope(s) with respect to variations in the real interest rate, holding constant the remaining exogenous factors. Figures 2 to 5 are constructed under the assumption that this property holds.

Figure 1 shows the similarities and differences between our model and the conventional two-period model, focusing exclusively on three variables: investment, saving, and the real interest rate.

**Figure 1.** *Determination of the real interest rate (in the absence of default risk) in the conventional model: "closed" economy ($M = X$) versus open economy*

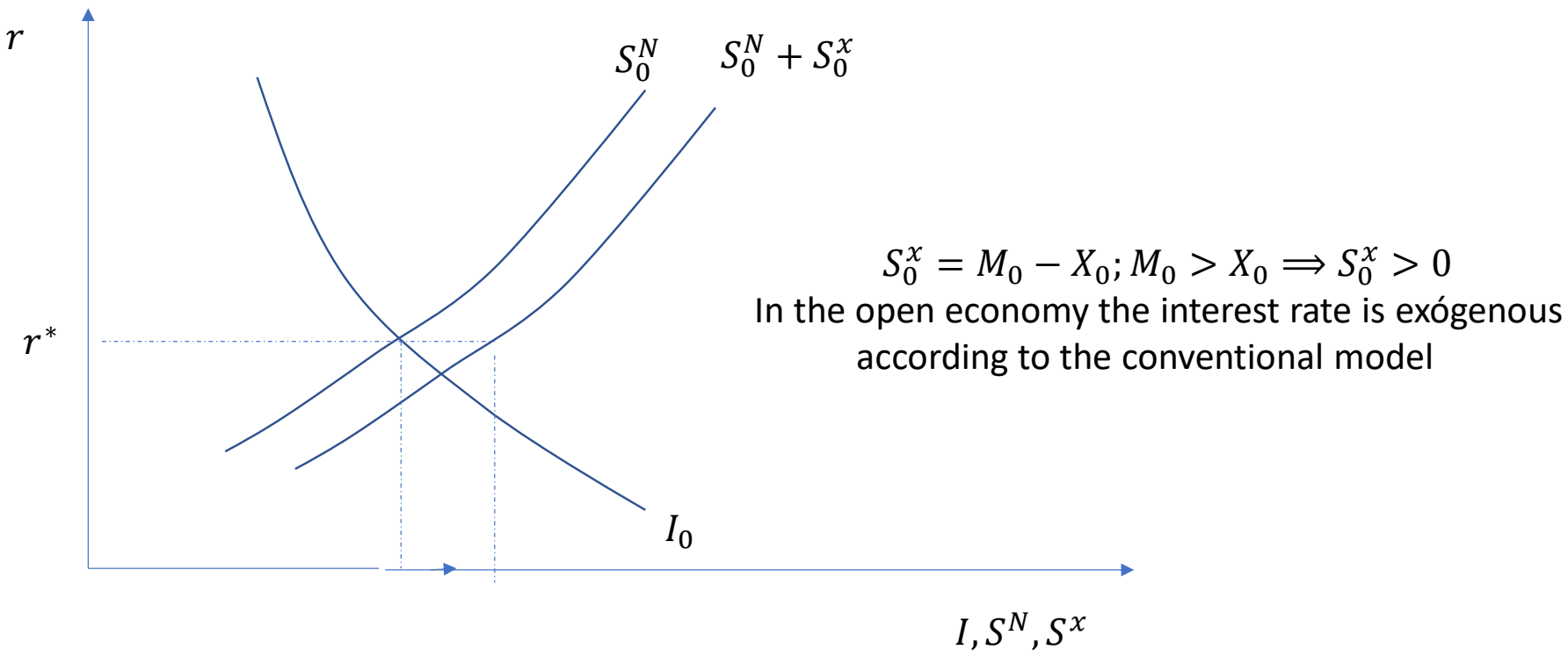

### *Some results of comparative static exercises*

Figures 2 to 5 show the effects of increases in $\gamma, \theta, \rho$ and $A_1$ (future multifactor productivity). These effects are captured through comparative statics exercises using a numerical version of the model presented in Section 3.3 (see Annex). In each exercise, the value of each one of these factors is increased by 15%, holding constant the values of the remaining parameters and exogenous variables (that is, maintaining them at their baseline levels). The results obtained by comparing each scenario with the baseline scenario confirm the importance of the relationships between the factors generating changes in the external trade balance or in the labor market and those inducing variations in investment, national saving, real interest rate and real wage.

**Figure 2.** *Comparison with the baseline scenario. Effects of increases in $\gamma$: present saving ($S_0^N$) increases and future saving ($S_1^x$) decreases but its magnitudes do not change so much, so the external deficit is practically the same, with no change in the real interest rate. Therefore, the curve $S_0^N + S_1^x$ does not shift, and investment remains unchanged.*

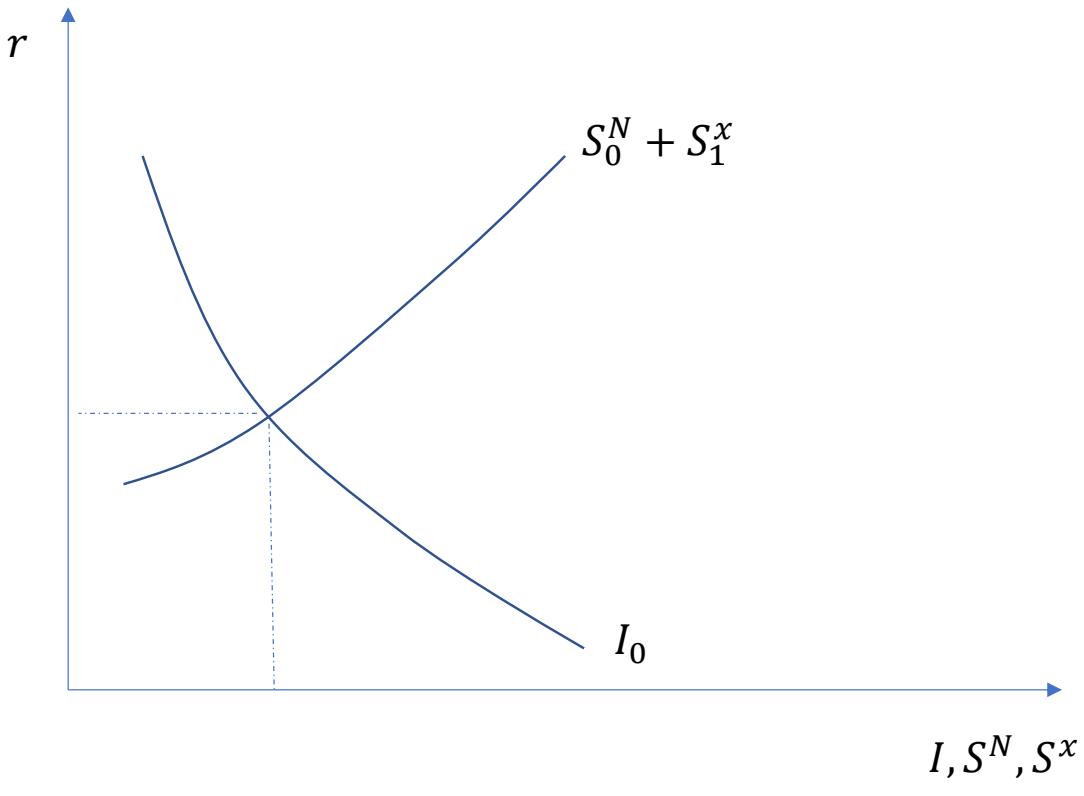

**Figure 3.** *Comparison with the baseline scenario. The increase in $\theta$ (decrease in the wage elasticity of labor supply) has several effects: a leftward shift of the savings curve and an increase in the interest rate; this shifts the labor supply curve to the right and reduces the wage; therefore, present output increases.*

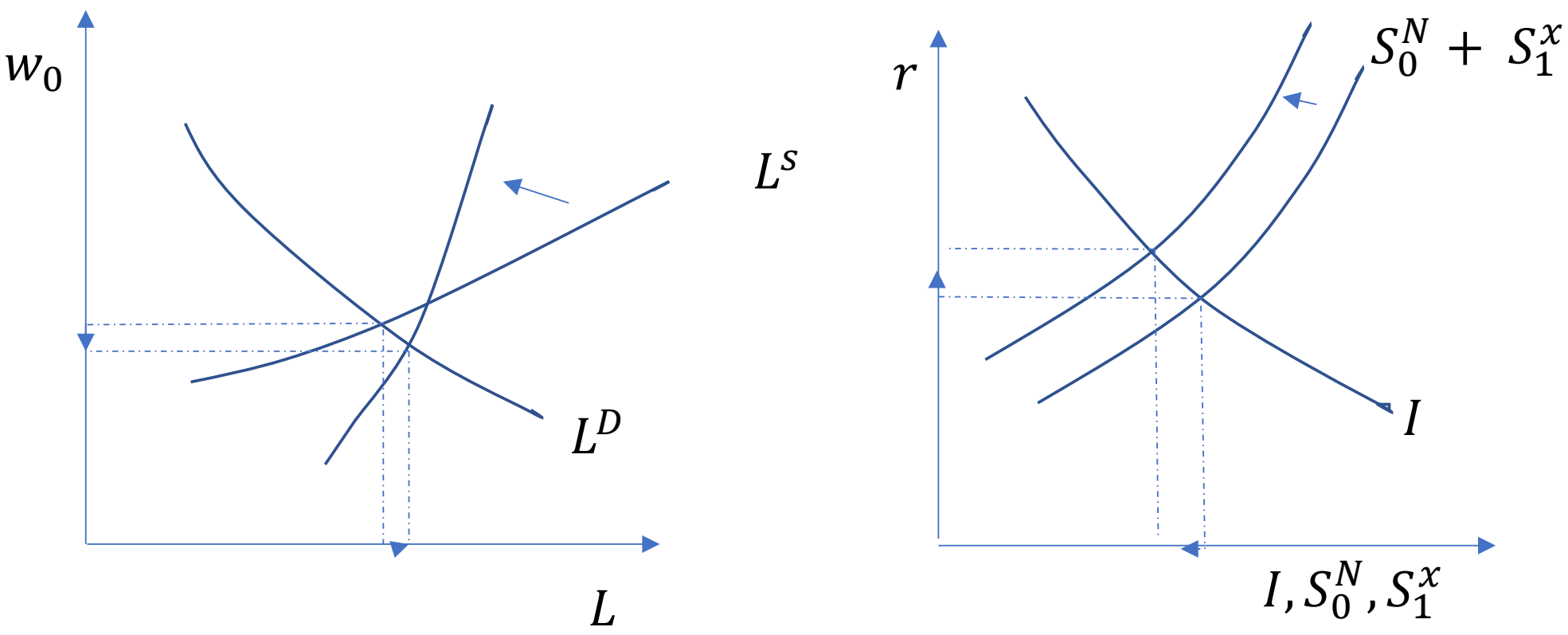

**Figure 4.** *Comparison with the baseline scenario. Effects of an increase in ρ: the interest rate, labor supply (the number of hours of work offered) and present output increase; the interest rate increases and, therefore, investment and present consumption fall; the wage and present external deficit decrease.*

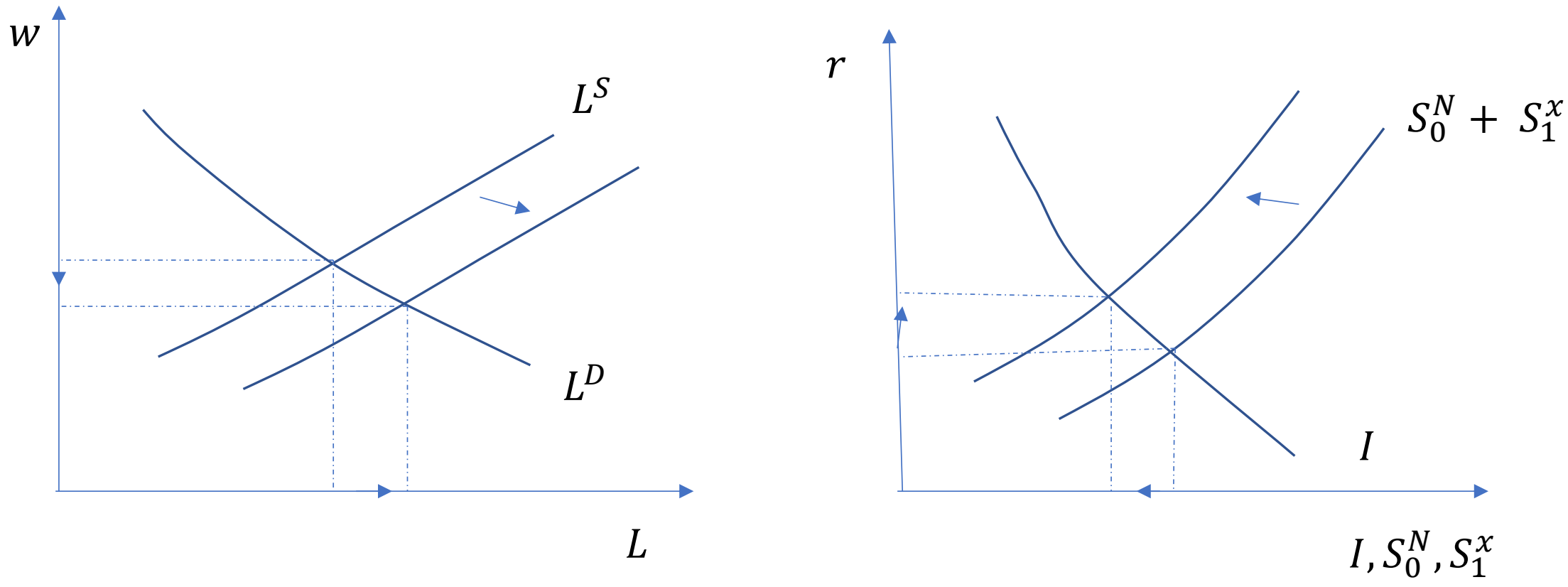

**Figure 5.** *Comparison with the baseline scenario. Effects of an increase in future multifactor productivity: the interest rate increases, but so do investment, consumption and wage. Labor supply decreases (due to the expectation of higher future income) and current output falls. The current external deficit increases.*

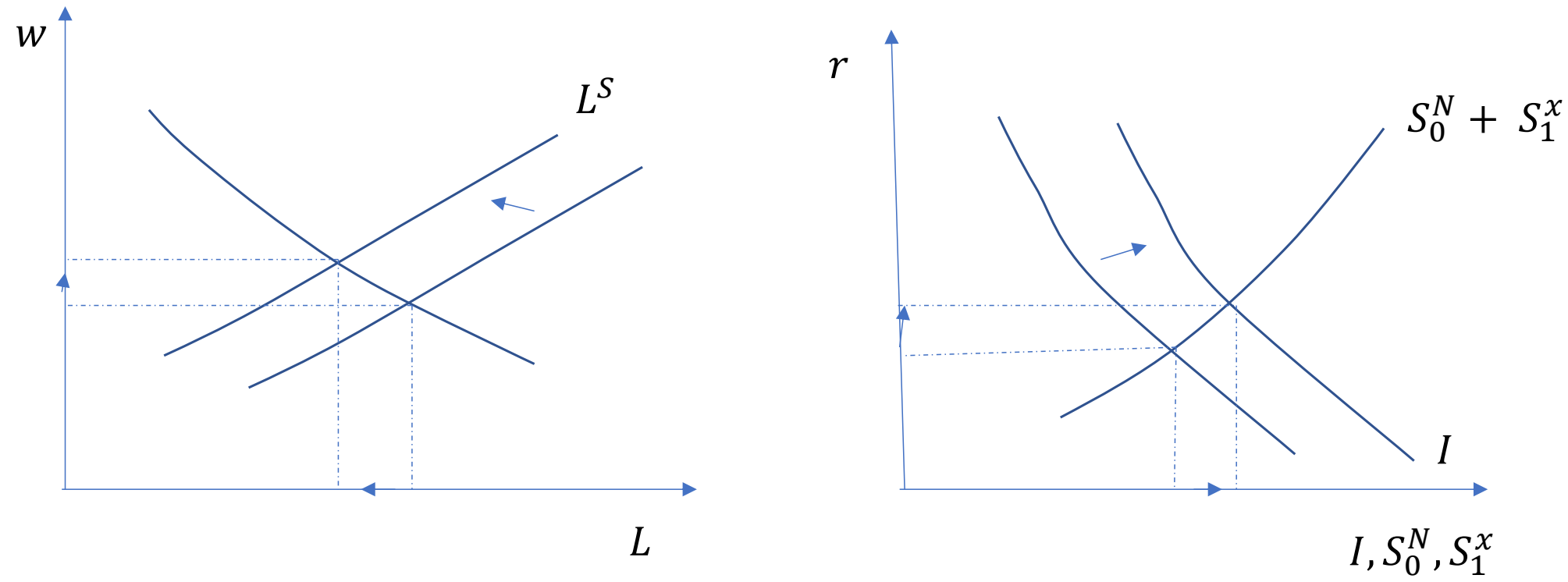

### *6. Summary and Conclusions*

This paper can be summarized as an argument in support of the following hypothesis: 1) The attempts to establish a substantive distinction between large and small open economies lack a theoretical foundation. Indeed, theory does not predict that large open economies are immune to default risk, nor that small ones are inherently risky in this regard. Nor does it predict that the basic features of the economic structure and dynamics governing the determination of the real interest rate in a large open economy differ from those of a small open economy. Therefore, if one abstracts from default´s risk, and if differences are observed across real interest rates in different open economies, the explanation for those differences must lie in differences in structural or dynamics characteristics of the economies themselves, independently of their size, as well as in the preferences of their agents. 2) It is reasonable to assume that the agents trading loanable funds agree on the definitive future moment when the debt incurred today (by the local economy or by the rest of the world) must be fully cancelled; That is, they agree on the duration of the future period related to the negotiation, and so they must accept that beyond that moment there is nothing more: that they must assume that "the world ceases to exist": In other words, there is no possibility of postponing commitments or forgiving obligations.

The main results from the comparative statics exercises based on numerical versions of the model can be summarized as follows[7]:

The determinants of the real interest rate are several, but two stand out in terms of their impact (see Table A-3, Annex):

i. The parameter $\rho$ (the subjective discount rate of future utility). The higher the value of this parameter, the higher the real interest rate. In this case, the increase in the real interest rate is sufficiently strong to reduce investment by an amount that, in absolute value, far exceeds an eventual increase in present consumption and, therefore, reduces the external trade deficit (or generates or increases a surplus in the trade balance). The increase in the interest rate does not reduce consumption because the intertemporal elasticity of substitution of consumption is low (= $1/\gamma$ = 0.83), and what raises the interest rate is the increase in the discount rate of future utility, which, in itself, has a positive effect on consumption.
ii. The exogenous variable $A_1$, which represents the level of multifactor productivity in the future period as anticipated in the present. An increase in this variable has a significant positive effect on the real interest rate. The current trade deficit increases for the following reasons: labor supply decreases due to the expectation of higher future income, leading to a fall in current output and an increase in current and future wages; furthermore, investment and current consumption increase because of a such expectation about $A_1$.

[7] Numerical simulations were performed for four alternative scenarios to the baseline scenario. In each simulation, only one exogenous factor (one of the following: γ, θ, ρ,$A_1$) was increased by 15% with respect to its value in the baseline scenario. A scenario involving an increase in parameter $\alpha$ was not constructed because the results would be counterintuitive for the following reason: in this model, as already noted, $\alpha$ is less than 1 and $\delta$ is equal to 1 (and if it were not equal to 1 but rather lower, it would in any case have to be close to 1, since the future period must have a duration of more than 10 years, and an annual depreciation rate of approximately 4.4% is reasonable); therefore, the ratio $\alpha/(\delta+r)$is necessarily less than 1 in a two-period model (both periods being finite), so that the optimal level of capital and, consequently, investment decrease in such a model with any increase in $\alpha$. Indeed, as $\alpha$ increases, the expression $(\alpha/(\delta+r))^{1/(1-\alpha)}$decreases when $\alpha<\delta+r$; see equation (6). By contrast, in models with annual, quarterly, or monthly frequency, any increase in $\alpha$ leads to an increase in the optimal capital stock, because in those models $\alpha>\delta+r$.

## Annex. Comparative Statics: Numerical Exercises

**Table A – 1**

**Parameters (Baseline Escenary)**

| $\alpha$ | $\gamma$ | $\delta$ | $\theta$ | $\rho$ | $l_0^{max}$ | $l_1^{max}$ |
|---|---|---|---|---|---|---|
| 0.5 | 1.2 | 1 | 9 | 0.5 | 35,000* | 29,440 hours* |

*Years/period: 16

$\delta$/year= 0.044; $\rho$/year = 0.026

**Table A – 2**

**Exogenous Variables (Baseline Escenary)**

| $A_0$ | $A_1$ | $N_0$ | $N_1$ | $K_0$ | $tax_0$ | $G_0$ | $G_1$ | $l_0^{max}$ | $l_1$ |
|---|---|---|---|---|---|---|---|---|---|
| 1 | 1 | 10 | 10 | 31,756 | 0 | 0 | 0 | 35,000 | 29,440 |

**Table A – 3**

| | Parameters | | | | |
|---|---|---|---|---|---|
| | **Baseline Escenary** | **Esc. II** | **Esc. III** | **Esc. IV** | **Esc. V** |
| $\alpha$ | 0.5 | 0.5 | 0.5 | 0.5 | 0.5 |
| $\delta$ | 1 | 1 | 1 | 1 | 1 |
| $\gamma$ | 1.2 | 1.38 | 1.2 | 1.2 | 1.2 |
| θ | 9 | 9 | 10.35 | 9 | 9 |
| ρ | 0.5 | 0.5 | 0.5 | 0.575 | 0.5 |
| $A_1$ | 1 | 1 | 1 | 1 | 1.15 |
| **Main Results of Comparative Static Exercises** | | | | | |
| $X0 - M0$ | -14948.74 | -14908.06 | -14812.42 | -11359.92 | -21991.62 |
| r | 0.4821 | 0.4821 | 0.4839 | 0.5560 | 0.4979 |
| Per year r | 0.0249 | 0.0249 | 0.0250 | 0.0280 | 0.0256 |
| Inv 0 | 33504.96 | 33504.96 | 33424.65 | 30397.05 | 37722.37 |
| Y 1 | 99316.97 | 99316.97 | 99197.87 | 94598.58 | 113010.11 |
| Y 0 | 96492.12 | 96492.12 | 96527.31 | 96739.03 | 95891.88 |
| $l_0$ | 29320.00 | 29320.00 | 29341.39 | 29470.25 | 28956.36 |
| C 0 | 77935.89 | 77895.21 | 77915.08 | 77701.90 | 80161.13 |
| C 1 | 77161.10 | 77221.39 | 77217.68 | 76921.99 | 80068.45 |
| Inv 0 / Y 0 | 0.3472 | 0.3472 | 0.3463 | 0.3142 | 0.3934 |
| C 0 / Y 0 | 0.8077 | 0.8073 | 0.8072 | 0.8032 | 0.8360 |
| w0/w1/(1+r) | 1.4459 | 1.4459 | 1.4488 | 1.5896 | 1.2923 |
| (X0-M0)/Y0 | -0.1549 | -0.1545 | -0.1535 | -0.1174 | -0.2293 |
| w0 | 0.1646 | 0.1646 | 0.1645 | 0.1641 | 0.1656 |
| w0/r | 0.3413 | 0.3413 | 0.3399 | 0.2952 | 0.3325 |
| w1 | 0.1687 | 0.1687 | 0.1685 | 0.1607 | 0.1919 |

**Note:** $per\ year\ r\ =\ (1+r)^{\frac{1}{16}} - 1$. That is, it is assumed that 16 years pass between the time of granting a loan (for example: at an intermediate point in the present period) and the time of its cancellation (at an intermediate point in the future period).